--------------------------------------------------------------
\documentstyle[eqsecnum,aps]{revtex}
\begin{document}

\title{The modified Seiberg-Witten monopole
equations \\
and their exact solutions}

\author{Zhong-Qi Ma \thanks{Electronic address: MAZQ@BEPC3.IHEP.AC.CN}}

\address{Institute of High Energy Physics, 
P.O.Box 918(4), Beijing 100039, The People's Republic of China}

\author{Bo-Yuan Hou}

\address{Graduate School, Chinese Academy of Sciences, 
Beijing 100039, The People's Republic of China}

%P. O. Box 3908, Beijing 100039, The People's Republic of China}

%\date{}

\maketitle

\vspace{4mm}

\begin{abstract}
The modified Seiberg-Witten monopole equations are presented
in this letter. These equations have analytic solutions in 
the whole $1+3$ Minkowski space with finite energy. The 
physical meaning of the equations and solutions
are discussed here.
\end{abstract}

\vspace{8mm}
Recently, Witten \cite{wit} developed an elegant dual approach, that 
much simplified the Donaldson theory \cite{don} of four-manifolds. 
This approach starts from the Seiberg-Witten monopole
equations \cite{sw}
$$\displaystyle \sum_{\mu=1}^{4}~\gamma_{\mu} D_{\mu} \psi({\bf r})
=\displaystyle \sum_{\mu=1}^{4}~\gamma_{\mu} \left(\partial_{\mu}
-ieW_{\mu}\right) \psi({\bf r})=0, \eqno (1) $$
$$F_{\mu \nu}^{+}({\bf r})= - \displaystyle {i \over 2} 
\overline{\psi}({\bf r})
\gamma_{\mu \nu}^{+} \psi({\bf r}). \eqno (2) $$

Witten noted that in the flat space the Seiberg-Witten equations
admit no $L^{2}$ solutions. However, Freund \cite{fre} found
a simple $non-L^{2}$ solution, based on a fermion moving
in a U(1) Dirac monopole field \cite{di}. This solution has 
singularity at the origin, and a singular string from the
monopole that can be removed by the concept of section \cite{wy}.

It is reasonable to extend Freund's solution by considering the 
solution based on a fermion moving in the smooth 't Hooft-Polyakov 
monopole field \cite{th,ps}. This solution is analytic in the 
whole space with finite energy such that it could not satisfy 
the Seiberg-Witten equations, but the modified ones. In other 
words, the second Seiberg-Witten equation (2) has to be modified 
to meet the SU(2) monopole solution.

As pointed out by Freund \cite{fre} that the spinor in his 
solution satisfies the Weyl-Dirac equation, and the spinor field
constructs the Coulomb field. It is our starting point that these 
important properties should be kept in the modified equations 
and the solutions. 

Prasad-Sommerfield \cite{ps} found the analytic SU(2) monopole 
solution with finite energy. If one changes the Higgs field into 
the fourth component of the gauge potential, an analytic, self-dual, 
static and spherically symmetric SU(2) monopole solution without 
external source can be obtained uniquely \cite{hou1}:
$${\bf W}=\displaystyle {\hat{r} \wedge {\bf T} \over er}
\left(1-r\phi(r) \right),~~~~~W_{4}=\displaystyle {iG(r) \over e}
{\bf T} \cdot \hat{r}, $$
$$\phi(r)=\displaystyle {\beta \over \sinh (\beta r)}\sim
\left\{\begin{array}{ll} r^{-1} &{\rm when}~~r \longrightarrow 0 \\
O(e^{-\beta r}) &{\rm when}~~ r \longrightarrow \infty \end{array}
\right. , $$
$$iG(r)=r^{-1}-\beta \coth (\beta r) \sim
\left\{\begin{array}{ll} O(r) &{\rm when}~~r \longrightarrow 0 \\
-\beta &{\rm when}~~ r \longrightarrow \infty \end{array}
\right. , \eqno (3) $$

\noindent
where $e$ is the unit electric charge, $\hat{r}$ is the unit
radial vector, $\beta$ is a constant with positive real part,
and ${\bf T}$ are the SU(2) generators in the isospin space.

The gauge field 
$$G_{\mu \nu}=\partial_{\mu} W_{\nu}-\partial_{\nu} W_{\mu}-
ie\left(W_{\mu}W_{\nu}-W_{\nu}W_{\mu}\right) \eqno (4) $$

\noindent
is self-dual such that the magnetic field $B_{k}=\epsilon_{ijk}
G_{ij}$ is equal to the electric field $-iE_{k}=G_{k4}$:
$${\bf B}=-i{\bf E}=B_{\parallel}(r) \hat{r}({\bf T}\cdot \hat{r})
 + B_{\perp}(r) \left({\bf T}-\hat{r}({\bf T}\cdot \hat{r}) \right), $$
$$B_{\parallel}(r)=- \displaystyle {1 \over er^{2} } + \displaystyle 
{\beta^{2} \over e \sinh ^{2}(\beta r)} \sim
\left\{\begin{array}{ll} -\beta^{2}/(3e^{2}) &{\rm when}~~
r \longrightarrow 0 \\
-1/(er^{2}) &{\rm when}~~ r \longrightarrow \infty \end{array}
\right. , $$
$$B_{\perp}(r)=\displaystyle {\beta \over er\sinh(\beta r) }
-\displaystyle {\beta^{2} \coth (\beta r) \over e\sinh (\beta r) }
\sim \left\{\begin{array}{ll} -\beta^{2}/(3e^{2}) &{\rm when}~~
r \longrightarrow 0 \\
O(e^{-\beta r}) &{\rm when}~~ r \longrightarrow \infty \end{array}
\right. . \eqno (5) $$

\noindent
Note that $B_{\parallel}(r)$ and $B_{\perp}(r)$ tend to the 
same constant when $r$ goes to zero so that it cancels to 
each other and ${\bf B}$ is single-valued at the origin. 

There is a standard technique \cite{ja,br,mo} 
to represent two SU(2) spins in a vector representation.
We sketch the idea in the following.

Transform the spinor part of magnetic field ${\bf B}$ into
$2\times 2$ matrix:
$$\sigma \cdot {\bf B}=\displaystyle \sum_{k=1}^{3}
\displaystyle \sum_{\tau=-1}^{1}~~\sigma^{k}B_{k}^{~\tau} T_{\tau},~~~~~
B({\bf r})_{ab}^{~~\tau} \equiv \displaystyle \sum_{k=1}^{3}~
\sigma^{k}_{ab}B_{k}^{~\tau}, \eqno (6) $$

\noindent
where $\sigma^{k}$ is the Pauli matrix with two spinor subscripts
$a$ and $b=\pm 1/2$, and $T_{\tau}$ is
another basis of isospin generator with subscript $\tau=1$, 
0, and $-1$,  
$$T_{1}=-2^{-1/2}\left(T_{x}+iT_{y}\right),~~~~
T_{0}=T_{z},~~~~T_{-1}=2^{-1/2}\left(T_{x}-iT_{y}\right). \eqno (7) $$

\noindent
Now, removing the basis $T_{\tau}$, we can rewrite the self-dual 
gauge field $B({\bf r})$ as a direct product of a $2\times 2$ 
matrix and a $3 \times 1$ matrix:
$$B({\bf r})=B_{\parallel}(r)\left(\begin{array}{cc} \cos \theta & \sin \theta 
e^{-i\varphi} \\ \sin \theta e^{i\varphi} & -\cos \theta  
\end{array}\right)\otimes \left(\begin{array}{c} -2^{-1/2}\sin \theta 
e^{-i\varphi} \\ \cos \theta \\ 2^{-1/2} \sin \theta e^{i\varphi} 
\end{array}\right)~~~~~~~~~~~~~~~~~~~~~~~~~~~~~~~~~~~~~~~~~~~~~~~~~~~$$
$$+B_{\perp}(r)\left\{\left(\begin{array}{cc} -\sin \theta & \cos \theta 
e^{-i\varphi} \\ \cos \theta e^{i\varphi} & \sin \theta  
\end{array}\right)\otimes \left(\begin{array}{c} -2^{-1/2}\cos \theta 
e^{-i\varphi} \\ -\sin \theta \\ 2^{-1/2} \cos \theta e^{i\varphi} 
\end{array}\right)+
\left(\begin{array}{cc} 0 & -i e^{-i\varphi} \\ i e^{i\varphi} 
& 0 \end{array}\right)\otimes \left(\begin{array}{c} i2^{-1/2} e^{-i\varphi} \\ 
0 \\ i2^{-1/2} e^{i\varphi} \end{array}\right) \right\}. \eqno (8) $$

\noindent
The second column of the spinor part is nothing but the complex 
conjugate of the first column:
$$B({\bf r})_{a(-1/2)}^{~~\tau}=\displaystyle \sum_{c=\pm 1/2} 
\sum_{\lambda=-1}^{1}~d^{1/2}_{ac}(\pi)d^{1}_{\tau \lambda}(\pi)
\left\{B({\bf r})_{c(1/2)}^{~~\lambda}\right\}^{*},~~~~~
d^{j}_{ab}(\pi)=(-1)^{j+a} \delta_{a(-b)},  \eqno (9) $$

\noindent
where the combinative coefficients have to be included due
to the similarity transformations $d^{j}(\pi)$ between two 
equivalent representations $D^{j}(SU(2))$ and $D^{j}(SU(2))^{*}$. 
Hence, removing the second column of the spinor part, we define 
self-dual field $G^{+}({\bf r})$ that is a direct product of 
two-component spinor and three-component isospinor:
$$G^{+}({\bf r})_{a}^{~\tau}=B({\bf r})_{a(1/2)}^{~~~\tau}, $$
$$G^{+}({\bf r})=B_{\parallel}(r)\left(\begin{array}{c} 
\cos \theta \\ \sin \theta e^{i\varphi} \end{array}\right)\otimes 
\left(\begin{array}{c} -2^{-1/2}\sin \theta 
e^{-i\varphi} \\ \cos \theta \\ 2^{-1/2} \sin \theta e^{i\varphi} 
\end{array}\right)~~~~~~~~~~~~~~~~~~~~~~~~~~~~~~~~~~~~~~~~~~~~~~~~~~~$$
$$+B_{\perp}(r)\left\{\left(\begin{array}{c} -\sin \theta \\ 
\cos \theta e^{i\varphi} \end{array}\right)\otimes 
\left(\begin{array}{c} -2^{-1/2}\cos \theta 
e^{-i\varphi} \\ -\sin \theta \\ 2^{-1/2} \cos \theta e^{i\varphi} 
\end{array}\right)+
\left(\begin{array}{c} 0 \\ i e^{i\varphi} \end{array} \right)\otimes 
\left(\begin{array}{c} i2^{-1/2} e^{-i\varphi} \\ 
0 \\ i2^{-1/2} e^{i\varphi} \end{array}\right) \right\}. \eqno (10) $$

On the other hand, let $\psi({\bf r})$ be a spinor field with 
spin 1/2 and isospin 1, and satisfy the Weyl-Dirac equation (1). 
By making use of the following $\gamma$ matrices:
$$\vec{\gamma}=\left(\begin{array}{cc} 0 & -i\vec{\sigma} \\ 
i \vec{\sigma} & 0 \end{array}\right),~~~~\gamma_{4}
=\left(\begin{array}{cc} 0 & {\bf 1} \\ 
{\bf 1} & 0 \end{array}\right),~~~~\gamma_{5}=
\left(\begin{array}{cc} {\bf 1} & 0  \\ 
0& {\bf -1} \end{array}\right), \eqno (11) $$

\noindent
we may introduce the two-component spinor fields
with isospin 1:
$$u^{\pm}({\bf r})\sim \displaystyle {1 \over 2}\left(1\pm \gamma_{5}
\right)\psi({\bf r}),~~~~~\psi({\bf r})=\left(\begin{array}{c}
u^{+}({\bf r}) \\ u^{-}({\bf r}) \end{array} \right). \eqno (12) $$

\noindent
Now, we modify the second Seiberg-Witten monopole
equation (2) as follows:
$$G^{+}({\bf r})_{a}^{~\tau}=u^{-}({\bf r})_{a}^{~\tau}, \eqno (13) $$

\noindent
where the subscripts $a=\pm 1/2$, and $\tau=1$, 0, $-1$. The main 
change from Eq.(2) to Eq.(13) is that the bilinear form of $\psi$ 
in Eq.(2) becomes the linear form in Eq.(13). It has a great merit 
of this change that the modified equations have analytic solutions 
in the whole Minkowski space with finite energy.

Twenty years ago the spinor solutions with isospin 1 of 
the Weyl-Dirac equation (1) were obtained \cite{mo,hou2,r} where 
the spinor field moves in an analytic, self-dual, static and spherically 
symmetric SU(2) monopole field without external source. Now, we 
sketch the calculation as follows.

Equation (1) is spherically symmetric so that the general
angular momentum ${\bf J}$ is conserved:
$${\bf J}={\bf L} +{\bf S} +{\bf T},~~~~~~{\bf S}=\vec{\sigma}/2. 
\eqno (14) $$

\noindent
The two-component spinors $u^{\pm}_{jm}({\bf r})$ can be expanded by
the common eigenfunctions of $J^{2}$, $J_{z}$,
$S^{2}$, $S\cdot \hat{r}$, $T^{2}$ and $T\cdot \hat{r}$:
$$u^{\pm}_{jm}({\bf r})_{a}^{~\tau}
=\displaystyle \sum_{b=\pm 1/2} \sum_{\lambda=-1}^{1}~
f^{\pm}_{jmb\lambda}(r)\eta^{j}_{mb\lambda}(\hat{r})_{a}^{~\tau}, \eqno (15)$$
$$\eta^{j}_{mb\lambda}(\hat{r})_{a}^{~\tau}=\left(\displaystyle 
{2j+1 \over 4\pi}\right)^{1/2}D^{j}_{m(b+\lambda)}(\varphi,\theta,0)^{*}
D^{1/2}_{ab}(\varphi,\theta,0)D^{1}_{\tau \lambda}(\varphi,\theta,0),
\eqno (16) $$

\noindent
where $\theta$ and $\varphi$ are the polar and azimuthal angles
of $\hat{r}$, the spinor subscript $a$ runs over $\pm 1/2$ and 
the isospinor subscript $\tau$ runs over $1$, $0$, and $-1$. 
Expressing the spinor and the isospinor
as a column matrix, respectively, we may rewrite
$\eta^{j}_{mb\lambda}(\hat{r})$ as a direct product of a $2\times 1$
matrix and a $3\times 1$ matrix, satisfying
$$J^{2}\eta^{j}_{mb\lambda}(\hat{r})
=j(j+1)\eta^{j}_{mb\lambda}(\hat{r}),~~~~
S^{2}\eta^{j}_{mb\lambda}(\hat{r})
=(3/4)\eta^{j}_{mb\lambda}(\hat{r}),~~~~
T^{2}\eta^{j}_{mb\lambda}(\hat{r})
=2\eta^{j}_{mb\lambda}(\hat{r}), $$
$$J_{z}\eta^{j}_{mb\lambda}(\hat{r})
=m\eta^{j}_{mb\lambda}(\hat{r}),~~~~
\left(S\cdot \hat{r}\right)\eta^{j}_{mb\lambda}(\hat{r})
=b\eta^{j}_{mb\lambda}(\hat{r}),~~~~
\left(T\cdot \hat{r}\right)\eta^{j}_{mb\lambda}(\hat{r})
=\lambda \eta^{j}_{mb\lambda}(\hat{r}). \eqno (17) $$

It is easy to show the following formulas:
$$\vec{\sigma} \cdot \vec{D}=\left(\vec{\sigma} \cdot \hat{r}
\right) \left(\displaystyle {\partial \over \partial r} + \displaystyle 
{1 \over r} \right) -r^{-1} \left(\vec{\sigma} \cdot \hat{r}
\right) K+i\phi(r) \vec{\sigma}\cdot \left(\hat{r} \wedge
{\bf T} \right), $$
$$K=\vec{\sigma} \cdot \left\{ {\bf r} \wedge
\left(-i\bigtriangledown -{\bf r} \wedge {\bf T}/r^{2} \right)\right\}
+1, \eqno (18) $$
$$\vec{\sigma}\cdot \left(\hat{r}\wedge {\bf T}\right)
\eta^{j}_{mb\lambda}(\hat{r})
=i2bA_{b\lambda}
\eta^{j}_{m(-b)(\lambda+2b)}(\hat{r}),~~~~
A_{b\lambda}=\left\{(9/4)-(b+\lambda)^{2}\right\}^{1/2}$$
$$K\eta^{j}_{mb\lambda}(\hat{r})
= K_{\lambda}\eta^{j}_{m(-b)\lambda}(\hat{r}),~~~~~~
K_{\lambda}=\left\{(j+1/2)^{2}-\lambda^{2}\right\}^{1/2}. \eqno (19) $$

\noindent
Hence, equation (1) becomes
$$\left(\displaystyle {\partial \over \partial r}+\displaystyle 
{1 \over r} \right) f^{\pm}_{jmb\lambda}(r)-K_{\lambda}r^{-1}
f^{\pm}_{jm(-b)\lambda}(r)+A_{b\lambda}\phi(r) 
f^{\pm}_{jm(-b)(\lambda+2b)}(r)
=\pm iG(r) 2b\lambda f^{\pm}_{jmb\lambda}(r). \eqno (20) $$

\noindent
From the convergent condition, the only analytical solutions
are obtained for $j=1/2$ \cite{hou2}:
$$f^{+}_{jmb\lambda}(r)=0,$$
$$f^{-}_{(1/2)m(1/2)0}(r)=-f^{-}_{(1/2)m(-1/2)0}(r)=c_{m}B_{\parallel}(r),$$
$$f^{-}_{(1/2)m(1/2)(-1)}(r)=-f^{-}_{(1/2)m(-1/2)1}(r)
=c_{m}\sqrt{2} B_{\perp}(r). 
\eqno (21) $$

\noindent
where $c_{m}$ is an arbitrary constant due to the linear
property of Eq.(1), and $B_{\parallel}$ and $B_{\perp}$ were 
given in Eq.(5). The rest components $f^{-}_{jmb\lambda}(r)$
are vanishing.

In terms of the exact forms of $D^{j}_{mm'}(\alpha,\beta,\gamma)$
\cite{ro}:
$$D^{j}_{mm'}(\alpha,\beta,\gamma)
=\displaystyle \sum_{n}~\displaystyle {(-1)^{n}\left\{
(j+m)!(j-m)!(j+m')!(j-m')!\right\}^{1/2} \over
(j+m-n)!(j-m'-n)!n!(n-m+m')! } $$
$$~~~~\cdot~e^{-im\alpha}\left(\cos (\beta/2)\right)^{2j+m-m'-2n}
\left(\sin (\beta/2)\right)^{2n-m+m'}e^{-im'\gamma}. \eqno (22) $$

\noindent
we have
$$\eta^{1/2}_{(1/2)(1/2)0}(\hat{r})-\eta^{1/2}_{(1/2)(-1/2)0}(\hat{r})
=(2\pi)^{-1/2}\left(\begin{array}{c} \cos \theta \\ \sin \theta 
e^{i\varphi} \end{array}\right)\otimes 
\left(\begin{array}{c} -2^{-1/2}\sin \theta 
e^{-i\varphi} \\ \cos \theta \\ 2^{-1/2} \sin \theta e^{i\varphi} 
\end{array}\right),$$
$$\eta^{1/2}_{(-1/2)(1/2)0}(\hat{r})-\eta^{1/2}_{(-1/2)(-1/2)0}(\hat{r})
=(2\pi)^{-1/2}\left(\begin{array}{c} \sin \theta e^{-i\varphi} \\
-\cos \theta \end{array}\right)\otimes 
\left(\begin{array}{c} -2^{-1/2}\sin \theta 
e^{-i\varphi} \\ \cos \theta \\ 2^{-1/2} \sin \theta e^{i\varphi} 
\end{array}\right),$$
$$\eta^{1/2}_{(1/2)(1/2)(-1)}(\hat{r})-\eta^{1/2}_{(1/2)(-1/2)1}(\hat{r})
~~~~~~~~~~~~~~~~~~~~~~~~~~~~~~~~~~~~~~~~~~~~~~~~~~~~~~~~~~~~~~~~~~~~~~$$
$$=(4\pi)^{-1/2}\left\{\left(\begin{array}{c} -\sin \theta \\ \cos \theta 
e^{i\varphi} \end{array}\right)\otimes 
\left(\begin{array}{c} -2^{-1/2}\cos \theta 
e^{-i\varphi} \\ -\sin \theta \\ 2^{-1/2} \cos \theta e^{i\varphi} 
\end{array}\right)+ (4\pi)^{-1/2}
\left(\begin{array}{c} 0 \\ i e^{i\varphi} \end{array}\right)\otimes 
\left(\begin{array}{c} i2^{-1/2} e^{-i\varphi} \\ 
0 \\ i2^{-1/2} e^{i\varphi} \end{array}\right) \right\}, $$
$$\eta^{1/2}_{(-1/2)(1/2)(-1)}(\hat{r})-\eta^{1/2}_{(-1/2)(-1/2)1}(\hat{r})
~~~~~~~~~~~~~~~~~~~~~~~~~~~~~~~~~~~~~~~~~~~~~~~~~~~~~~~~~~~~~~~~~~~~~~$$
$$=(4\pi)^{-1/2}\left\{\left(\begin{array}{c} \cos \theta e^{-i\varphi} 
\\ \sin \theta \end{array}\right)\otimes 
\left(\begin{array}{c} -2^{-1/2}\cos \theta 
e^{-i\varphi} \\ -\sin \theta \\ 2^{-1/2} \cos \theta e^{i\varphi} 
\end{array}\right)+ (4\pi)^{-1/2}
\left(\begin{array}{c} -i e^{-i\varphi} \\ 0 \end{array}\right)\otimes 
\left(\begin{array}{c} i2^{-1/2} e^{-i\varphi} \\ 
0 \\ i2^{-1/2} e^{i\varphi} \end{array}\right) \right\}, \eqno (23) $$

If we choose the constant $c_{m}$ in (21) as follows:
$$c_{1/2}=c_{-1/2}=(2\pi)^{1/2}, \eqno (24) $$

\noindent
then, $u^{+}_{jmb\lambda}({\bf r})=0$, and 
$u^{-}_{(1/2)(1/2)b\lambda}({\bf r})$ just satisfies the modified 
Seiberg-Witten monopole equations (1) and (13). It is evident that 
$u^{-}_{(1/2)(1/2)}({\bf r})$ and 
$u^{-}_{(1/2)(-1/2)}({\bf r})$ satisfy a complex conjugate 
relation like Eq.(9):
$$u^{-}_{(1/2)(-1/2)}({\bf r})_{a}^{~\tau}=\displaystyle \sum_{c=\pm 1/2} 
\sum_{\lambda=-1}^{1}~d^{1/2}_{ac}(\pi)d^{1}_{\tau \lambda}(\pi)
\left(u^{-}_{(1/2)(1/2)}({\bf r})_{c}^{~~\lambda}\right)^{*}, \eqno (25) $$

In summary, the modified Seiberg-Witten equations consist of
Eq.(1) and Eq.(13). Equation (1) describes a spinor field with 
isospin 1 moving in an analytic, self-dual, static and spherically 
symmetric SU(2) monopole field without external source. And 
equation (13) shows that the spinor field relates with the gauge 
field directly. It is the reason why the gauge field still 
satisfies the Yang-Mills equation without external source. 

Since the gauge field is a hedgehog solution, its integral
on a closed spherical face is vanishing:
$$\int {\bf B}\cdot {\bf dS} =0. \eqno (25) $$

\noindent
But, 
$$\displaystyle {e \over 8\pi} Tr\left\{
\int {\bf B}\left({\bf T}\cdot \hat{r} \right)\cdot {\bf dS} \right\}
=er^{2} B_{\parallel}(r) \sim -1,~~~~~{\rm when}~~r 
\longrightarrow \infty. \eqno (26) $$

\noindent
This is the first Chern number of the gauge field, and the asymptotic
form shows that the total magnetic charge  is $-1/e$. Furthermore,
$$\left(4\pi r^{2}\right)^{-1}Tr\left\{
\int \displaystyle \sum_{\mu \nu \rho \sigma}~
\epsilon_{\mu \nu \rho \sigma} G_{\mu \nu} G_{\rho \sigma}
dS \right\}=2B_{\parallel}(r)^{2}+4B_{\perp}(r)^{2}. \eqno (27) $$

On the other hand, the spinor field $\psi_{jm}({\bf r})$ with
$j=1/2$ satisfies:
$$\left(4\pi r^{2}\right)^{-1} \int \psi_{jm}({\bf r})^{\dagger}
\left( \vec{\sigma} \cdot \hat{r} \right) \psi_{jm}({\bf r}) dS
=\left(4\pi r^{2}\right)^{-1} \int \psi_{jm}({\bf r})^{\dagger}
\left({\bf T}\cdot \hat{r} \right) \psi_{jm}({\bf r}) dS
=0. \eqno (28) $$
$$\left(4\pi r^{2}\right)^{-1} \int \psi_{jm}({\bf r})^{\dagger}
 \psi_{jm}({\bf r}) dS=(4\pi)\left(B_{\parallel}(r)^{2}
+2B_{\perp}(r)^{2}\right). \eqno (29) $$

\noindent
Therefore,
$$\left(16\pi^{2}\right)^{-1}~ Tr\left\{
\int \displaystyle \sum_{\mu \nu \rho \sigma}~
\epsilon_{\mu \nu \rho \sigma} G_{\mu \nu} G_{\rho \sigma}
dS \right\}=\left(32\pi^{3}\right)^{-1}\int \psi_{jm}({\bf r})^{\dagger}
 \psi_{jm}({\bf r}) dS. \eqno (30) $$

\noindent
where the integrand on the left-hand-side of Eq.(30) is nothing 
but the second Chern class.

The physical meaning of the modified Seiberg-Witten monopole
equations should be further explored, and new solutions will
be sought later.

\vspace{10mm}
{\bf Acknowledgments} The authors would like to thank Prof. Yu-Fen
Liu and Mr. Fu-Zhong Yang for useful discussions. 
This work was supported by the National
Natural Science Foundation of China and Grant No. LWTZ-1298 of
the Chinese Academy of Sciences.

%\vspace{-5mm}
%\vspace{10mm}


\begin{thebibliography}{99}
\bibitem{wit} E. Witten, Math. Res. Lett. {\bf 1}, 769 (1994).

\bibitem{don} S. Donaldson, J. Diff. Geom. {\bf 18}, 279 (1983); 
Topology {\bf 29}, 257 (1993).

\bibitem{sw} N. Seiberg and E. Witten, Nucl. Phys. {\bf B426},
19 (1994).

\bibitem{fre} P. G. O. Freund, J. Math. Phys. {\bf 36}, 2673 (1995).

\bibitem{di} P. A. M. Dirac, Phys. Rev. {\bf 74}, 817 (1948).

\bibitem{wy} T. T. Wu and C. N. Yang, Phys. Rev. {\bf D12}, 3845
(1975).

\bibitem{th} 't Hooft, Nucl. Phys. {\bf B79}, 276 (1974);
A. M. Polyakov, JETP Lett. {\bf 20}, 194 (1974).

\bibitem{ps} M. K. Prasad and C. M. Sommerfield, Phys. Rev.
Lett. {\bf 35}, 760 (1975).

\bibitem{hou1} Bo-Yu Hou and Bo-Yuan Hou, Phys. Ener. Fort.
Phys. Nucl. {\bf 3}, 255 (1979), English version: Chin. Phys. 
{\bf 1}, 624 (1981).

\bibitem{ja} R. Jackiw and C. Rebbi, Phys. Rev. {\bf D23}, 3398
(1976).

\bibitem{br} L. S. Brown, R. D. Carlitz and C. Lee, Phys. Rev.
{\bf D16}, 242 (1978).

\bibitem{mo} E. Mottola, Phys. Lett. {\bf 79B}, 242 (1978).

\bibitem{hou2} Bo-Yu Hou and Bo-Yuan Hou, Phys. Ener. Fort.
Phys. Nucl. {\bf 3}, 697 (1979).

\bibitem{r} P. Rossi, Phys. Report {\bf 86}, 317 (1982).

\bibitem{ro} M. E. Rose, {\it Elementary Theory of Angular
Momentum}, John Wiley \& Sons, Inc., 1957.

\end{thebibliography}
\end{document}